\documentclass[twocolumn]{openjournal}
\usepackage{lipsum}
\usepackage{xcolor}
\usepackage{textgreek}
\usepackage[utf8]{inputenc}
\usepackage[english]{babel}

\usepackage{hyperref}
\hypersetup{
    unicode, 
    colorlinks=true,
    linkcolor=linkcolor,
    citecolor=linkcolor,
    filecolor=linkcolor,
    urlcolor=linkcolor,
}
\usepackage{color,colortbl}
\definecolor{linkcolor}{rgb}{0.0,0.3,0.5}
\DeclareGraphicsExtensions{.bmp,.png,.jpg,.pdf}
\usepackage{verbatim}
\usepackage[normalem]{ulem}
\usepackage{orcidlink}
\usepackage{soul}
\usepackage{float}
\usepackage{algorithm}
\usepackage{algpseudocode}
\usepackage{amsmath}
\usepackage{nth}
\graphicspath{ {./figs/} }
\DeclareUnicodeCharacter{0306}{\u{}}
\begin{document}
\title{StarHash: unique, memorable, and deterministic names for astronomical objects}
\author{T. L. Killestein\orcidlink{0000-0002-0440-9597}}
\email{tom.killestein@gmail.com}
\affiliation{Department of Physics, University of Warwick, Gibbet Hill Road, Coventry CV4 7AL, UK}
\vspace{-5mm}
\begin{abstract}
    The naming of astronomical objects has represented among the most significant challenges in the record-keeping of the field since the very beginning. Long and unwieldy coordinate names, uninformative and ambiguous internal names, and the sheer volume of aliases accumulated for some of the most studied objects conspire to complicate our study of the celestial sphere.
    This paper proposes \textsc{StarHash}, a reproducible, open-source astronomical naming scheme based on the terrestrial concept of geohashing, but re-implemented from the ground up for the rigorous demands of astronomy. Every 3.2 arcsecond patch of sky now has three words associated with it, enabling the precise localisation of astronomical sources, and an easily communicable and memorable identifier. A carefully selected wordlist reduces ambiguity due to plurals and homophones, whilst the use of format-preserving encryption minimises residual spatial correlation in \textsc{StarHash}-derived identifiers. Pre-computed names for several existing catalogues are provided, alongside a Python reference implementation for validation and integration into databases, transient brokers, and other similar projects. Although not intended to be the final word in the naming of astronomical objects, \textsc{StarHash} humbly provides a memorable alternative to the status quo, and is intended to spark a discussion about this most foundational of issues in astronomy. 
\end{abstract}
\maketitle
\section{Introduction}
Astronomical object naming is a mess -- and has remained a mess for quite some time. Even in the very earliest catalogues, objects acquired multiple names, with Messier and Herschel cataloguing the same objects under different names in their respective catalogs~\citep{Messier1781,Herschel1864}, in spite of being aware of each others works. This tradition continued with the New General Catalog~\citep[NGC; ][]{Dreyer1888} and Index Catalog~\citep[IC; ][]{Dreyer1895}, which both introduced new identifiers for existing, well-known objects. 

The naming of variable stars in particular exemplifies the complexities of astronomical naming schemes, starting off with the Bayer~\citep{Bayer1603} designation (e.g. $\alpha$~Ori), and quickly descending into chaos with a series of arcane rules for the prefix adopted as the number of known variable stars grew. First, the letters R-Z, then RR-ZZ, then AA-QZ (but never including J!), then transitioning to V335 and higher numbers once the number of variable stars in a constellation exceeds the available letter combinations.

A common solution for object naming is the ``telephone number'' style identifiers, composed of the catalog identifier, followed by the right ascension and declination -- for example, WD 1145+017, SDSS J1038+4849, and 2MASS J11193254–1137466 AB. There is no consensus on the format nor precision on the coordinates of such names, and worse, typographical errors in these names could lead to small (arcsec) to large (entire degrees) on-sky distances, depending on which digit is affected. Long coordinate-based names for objects can make them unintelligible to those not in the field, and disambiguating these names based on the subtly-different astrometry of each survey is challenging for humans. Maintaining a list of cross-identifications is important to ensure objects can be recognised under whichever name they are searched for by.

A number of object registries~\citep[e.g. ][]{Helou1991,Wenger2000} exist to address this problem, associating astronomical objects with their coordinates and myriad alternative names, painstakingly collected from the academic literature -- a significant challenge given the accelerating rates of publication, and the rapidly growing volume of new objects discovered by new surveys. Many of the more significant objects have tens of cross-identifications -- for example, at the time of writing, 
{\raggedright
\texttt{11HUGS\,027,
1AXG\,J013351+3039,
1ES\,0131+303,
$\ldots$
UGC\,01117,
UZC\,J013351.1+303922,
WISEA\,J013350.89+303936.7},
}
better known as the Triangulum Galaxy has 65 distinct identifiers\footnote{\LaTeX was unable to render the full list in a sensible way, so a truncated version appears here.} on the NASA Extragalactic Database~\citep[NED; ][]{Helou1991}. Minimising the number of names under which an object is known is crucial, to reduce the potential for confusion, and to make the job of librarians of astronomy and curators of astronomical databases significantly easier.

The rapid surge in the numbers of astronomical time-domain surveys and their survey capacity over the past decade has pushed the issues with astronomical naming towards new levels of confusion. Each survey that detects a given transient event will derive their own unique ``internal name'', carrying information about the epoch of discovery, and which survey it came from, but carrying no additional meaning and being entirely incomparable between surveys. Until reporting, these names often remain proprietary and cannot trivially be cross-referenced. For example, the Asteroid Terrestrial-impact Last Alert System~\citep[ATLAS; ][]{Tonry2018} were first to report a given transient with the internal designation ATLAS24fsk, which was confirmed by the Gravitational-wave Optical Transient Observer~\citep[GOTO; ][]{Steeghs2022,Dyer2024} and given the name GOTO24aig. BlackGEM~\citep{Groot2024} and PanSTARRS~\citep{Chambers2016} added to the rapidly growing list of names with BGEM J111822.10-325015.1 and PS24brj, respectively. This event is better known as SN\,2024ggi, thanks to the existence of an international clearing house for transient discoveries, the Transient Name Server (TNS\footnote{\url{https://wis-tns.org}}).

The establishment of the International Astronomical Union TNS was a great leap forward in reducing ambiguity surrounding transient naming -- upon reporting, objects are assigned a definitive IAU name, and their coordinates are registered to unambiguously link the identifier with that particular event. This harmonised the naming of astrophysical transients, presented an alternative to internal survey names and the prior PSN J1234-style names used within the prior IAU CBAT schema, and brought much-needed order to the field. Standards, however, are only as good as their adoption level and community adherence, and internal survey names remain in use, including in manuscript titles. The issue of what to call a given object \emph{prior} to reporting remains, and that the name changes upon reporting introduces another step where the object's identity may be confused.

Both the TNS naming schema and internal survey names broadly follow a sequential pattern, starting with the year of discovery, then appending letters for each discovery made or alert issued: progressing from \texttt{A-Z}, then \texttt{aa-az}, \texttt{ba-bz}, and so on, yielding (possibly) familiar names such as SN\,1987A and AT\,2018cow/ATLAS18qqn. This scheme is simple and purely relies on the number of transient discoveries/candidates so far in a given year to generate the name -- thus also relies on a centralised authority to keep track. In the case of sequential names, even a simple single-character error can have dire consequences\footnote{Multiple private communications; correspondents requested anonymity due to the unfortunate nature of the incidents.} -- especially in the high-stakes, high time pressure environment of time-domain astronomy. Reducing this ambiguity frees up cognitive load, and minimises the chance of precious telescope time being triggered on the wrong object. The TNS is ostensibly for extragalactic transients, which excludes the possibility of other objects, for example variable stars, being expeditiously being assigned official names through this mechanism.

There is a more subtle issue in light of real-time surveys that has not yet been raised in the literature: given that survey strategies focus on observing contiguous patches of sky, objects discovered sequentially will likely be closer on the sky than one might expect from random chance -- especially in the case of deeper surveys, where multiple live transients may be found in the same discovery image. This has an slightly insidious side effect: single-character transcription errors could lead to misidentifications that are nearby enough to not appear initially suspicious. These plausibly-wrong errors have the potential to pass initial checks, and be especially demoralising when astronomers spot them.

With the advent of Rubin Observatory's Legacy Survey of Space and Time~\citep{Ivezic2019}, this problem will only grow worse. The increased discovery volume will generate a deluge of new objects in need of names, and the Rubin internal alert names are long integers, which are very prone to transcription errors when transmitted via the lossy medium of human communication. 

Bearing the discussion in this Section in mind, the ideal name for an astronomical object should be:
\begin{enumerate}
    \item Related to the on-sky location of the object in a well-defined and reproducible, for easy cross-referencing between surveys and facilities
    \item Resistant to corruption via errors in communication or input into systems, with small on-sky separations mapping to large differences between identifiers
    \item Memorable, easily communicable, and entirely distinct from existing object identifiers
\end{enumerate}
Whilst the adoption of yet another naming scheme for astronomical objects is sure to be met with some consternation, I here lay out a foundation for an deterministic, open-source, and memorable scheme for the naming of astronomical objects -- \emph{beyond} the traditional paradigm of sequential or telephone-number-style coordinate names, taking inspiration from terrestrial location-finding. This paper introduces the new \textsc{StarHash} standard for astronomical object naming, designed from the sky down to address issues with name confusion, and inject a scientifically-justifiable amount of whimsy into our shared quest to understand the Universe.

\section{The StarHash Algorithm} \label{sec:methods}
Following the geographical precedent established by \textsc{what3words}~\citep{what3words}, I extend the concept of ``geohashing'' to the celestial sphere. The idea that three words present a memorable and easily communicable identifier is not a new one -- finding broad utility in terrestrial contexts -- but astronomy has distinct requirements and thus necessitates starting from scratch. This section lays out the initial reference implementation of \textsc{StarHash}, and elaborates upon the core design decisions and philosophy behind the algorithm. 

The first steps of \textsc{StarHash} require partitioning the celestial sphere into discrete chunks, ideally each covering the same area/spatial extent. Whilst there are potentially many such schemes, \textsc{StarHash} leans on Hierarchical Equal Area isoLatitude Pixelisation \citep[HEALPix; ][]{Gorski2005}, as the \emph{de facto} standard for spatial queries/operations in astronomy. The reference implementation of \textsc{StarHash} uses a HEALPix grid with \textsc{nside}=$2^{16}$, and the \textsc{ring} pixel ordering, which yields an approximate pixel size of 3.2 arcsec, and $\approx51$ billion pixels. This specific \textsc{nside} parameter is chosen to approximately match the typical association radii used in practice by existing name servers, and be conservative given the square HEALPix pixels.

Given that the intention of HEALPix is to cluster spatially nearby pixels in sequential HEALPix indices, a random, reversible mapping must be applied to scramble any spatial correlation -- which would yield \textsc{StarHash} phrases that share common words. Prior works have leveraged linear congruential generators \citep[LCG; ][]{linearcongruentialgenerator} in the domain of geohashing, to some success. Although very fast, the quality of randomness generated by LCGs is highly sensitive to the chosen parameters. If a LCG with a long enough period is not used, locations close by in geographical distance may also end up close in the hash space, an issue reported on both in the literature~\citep[e.g. ][]{Arthur2023}, and wider media. 

To sidestep this issue, \textsc{StarHash} employs format-preserving encryption, specifically the FF3~\citep{ff3} cipher, as implemented in the \texttt{ff3} Python package\footnote{\url{https://pypi.org/project/ff3/}}. The unique property of format-preserving encryption, as the name implies, is that the plaintext and ciphertext maintain the same length and format -- preserving the integer type of the HEALPix index in the output, which hereafter is referred to as the encrypted index. After obtaining the HEALPix index via the \textsc{ang2pix} function, zero padding is applied to convert it to a fixed-length string -- set to be the length of the maximum HEALPix index on the grid used. This zero-padded index string is then encrypted using the FF3 algorithm, using hard-coded values for the encryption key (\texttt{starhash!}) and the required tweak~\citep[see][]{ff3} value (\texttt{opensource}). The encrypted index is then converted back to an integer in preparation for assigning words to the resultant hash space.

The use of symmetric encryption, a bijective mapping, ensures that no name collisions occur, and that every HEALPix index is guaranteed to have a unique name associated with it. The quality of randomness provided by encryption is step up from LCG implementations, which ensures that the degree of spatial correlation is markedly lower. As of 2021, FF3 has been found to be cryptographically insecure~\citep[e.g.][]{ff3_weakness,ff3_weakness2}, but given that the naming of astronomical objects requires zero guarantees on the security of the encryption, this is a minor detail. Indeed, both the key and tweak required are hard-coded into the reference implementation.

The choice of word list is an important consideration in the implementation of \textsc{StarHash}. It must have sufficient words to cover all pixel indices using three words, preferably multiple times for redundancy, whilst avoiding homophones, plurals, and other confusing words. As curating a wordlist is a challenging task, I opt instead for using the existing EFF Long Wordlist\footnote{\url{https://www.eff.org/dice}}, which contains 7,776 English words. This list is already optimised for the memorability of words, as well as removing profane and/or difficult to spell words. This is important for both broad adoption, as well as academic sensibility, and avoids prior issues with astronomical objects being assigned \emph{unfortunate} names~\citep[see][discovered by citizen scientists via Kilonova Seekers]{sn2025ass}. The word list is also augmented with a list of 226 more specialised words from the astronomical lexicon, compiled by the Author to boost the wordlist to a total of 7,971 words.

For a wordlist $W$ with $N_w$ members, arranged into $k$ words, the total number of permutations is $W^k$. The coverage factor, $C$ is defined:
$$C = \frac{{N_w}^k}{N_{pix}} = \frac{{N_w}^k}{12N_{side}^2}$$ where $N_{pix}$ is the number of HEALPix pixels in the grid. This must be greater than one to ensure full coverage of the hash space, but not much greater to ensure efficient use of memory. Some overprovisioning does provide advantages in terms of widening the ``invalid'' hash space.

The indices to the wordlist are obtained by repeated division of the encrypted index modulo $N_w$, or in other words the base-$N_w$ decomposition -- ensuring a one-to-one mapping between the word indices and encrypted index, easily reversed by summing each word index raised to the power of its' position. The final phrase is formed by concatenation of the $k$ words with hyphens. The forward pass is documented in pseudocode form in Algorithm~\ref{alg:forwardpass}.
\begin{algorithm}[H]
    \caption{\textsc{StarHash} coordinates $\rightarrow$ words}\label{alg:forwardpass}
    \begin{algorithmic}[1]
        \Require $(\alpha, \delta)$ sky coordinates, $W$ wordlist of length $N_w$, number of words $k$, HEALPix grid of order $N_\mathrm{side}$
        \Ensure A $k$-word StarHash string $s$
        \State $p \gets \Call{ang2pix}{\alpha, \delta, N_\mathrm{side}}$
        \State $p' \gets \textsc{FF3Encrypt}(\textsc{ZeroPad}(p))$ \Comment{Zero-pad to fixed width}
        \State $t \gets p'$
        \For{$i = 1$ to $k$} 
        \State $w_i \gets W[t \bmod N_w]$ \Comment{Extract words}
        \State $t \gets \lfloor t / N_w \rfloor$
        \EndFor
        \State \Return $s = w_1 \texttt{-} w_2 \texttt{-} \cdots \texttt{-} w_k$
    \end{algorithmic}
\end{algorithm}
The reverse pass runs in much the same way -- parsing the words and converting them back to word list indices, recovering the encrypted index via reversing the modulo division, and decrypting it. For completeness, this is documented in Algorithm~\ref{alg:reversepass}. 
In the Python reference implementation of \textsc{StarHash}, the round-trip property is automatically covered as part of the test suite.
\begin{algorithm}[H]
\caption{\textsc{StarHash} words $\rightarrow$ coordinates}\label{alg:reversepass}
\begin{algorithmic}[1]
\Require Name string $s$, wordlist $W$ of size $N_w$, number of words $k$, HEALPix grid of order $N_\mathrm{side}$
\Ensure $(\alpha, \delta)$ sky coordinates
\State Parse $s \to (w_1, w_2, \ldots, w_k)$
\For{$i = 1$ to $k$}
    \State $j_i \gets \mathrm{index}(w_i, W)$ \Comment{Extract word-list indices for the $i^\mathrm{th}$ word}
\EndFor
\State $p' \gets \sum_{i=1}^{k} j_i \cdot N_w^{\,i-1}$ \Comment{Recover encrypted index}
\State $p \gets \textsc{FF3Decrypt}(\textsc{ZeroPad}(p'))$
\State $(\alpha, \delta) \gets \textsc{pix2ang}(p, N_\mathrm{side})$
\State \Return $(\alpha, \delta)$
\end{algorithmic}
\end{algorithm}
For this reference implementation of \textsc{StarHash}, $k$ is set to three, following both the rule of threes and prior art. For the \textsc{nside}=$2^{16}$ HEALPix grid used, and the combined wordlist, the coverage factor is $\approx10$, more than adequate. With this crucial choice made, every 3.2 arcsec patch of sky now has three words uniquely associated with it, forming an informative, reversible, and distinctive identifier by which to name it. 

\section{Examples and evaluation} \label{sec:examples}
In this section, I provide a few examples of the usage of \textsc{StarHash} in practice - both from a perspective of validating the technical promises of the algorithm, along with illustrating the memorability of \textsc{StarHash} names in practice.

\begin{figure}
    \centering
    \includegraphics[width=\linewidth]{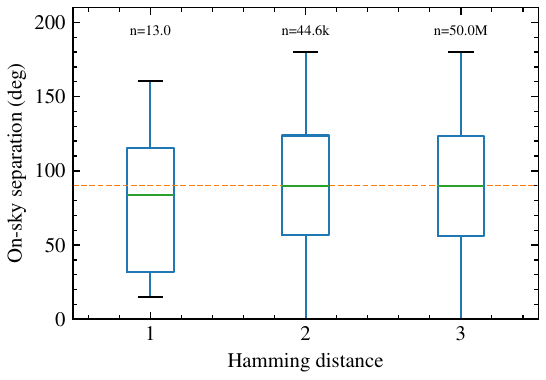}
    \caption{Box plot of Hamming distance of $\approx$50M unique coordinate pairs uniformly distributed across the sky and their on-sky distance. The box spans the \nth{1} and \nth{3} quantiles, with the whiskers showing 1.5$\times$ the inter-quartile range. The dashed orange line corresponds to the expectation of on-sky distance for the uniform sky distribution.}
    \label{fig:allsky_hamming}
\end{figure}

\begin{figure}
    \centering
    \includegraphics[width=\linewidth]{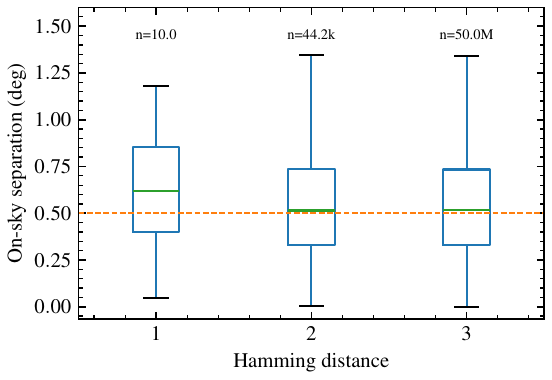}
    \caption{Box plot of Hamming distance of $\approx$50M coordinate pairs drawn from a Normal distribution of standard deviation 0.5 degrees. The boxes retain the same meaning as Figure~\ref{fig:allsky_hamming}, but in this case any results sharing the same HEALPix index have been removed, as an artefact of the sampling process.}.
    \label{fig:andromeda_hamming}
\end{figure}

In Section~\ref{sec:methods} it was asserted that the decision to use format-preserving encryption was driven by reducing the possibilities of spatial correlation arising from shared words between phrases. By comparing pairs of coordinates, their on-sky separation, and some relevant metric to assess distance in the word/code space, this can be empirically validated. The most relevant metric to consider is the Hamming distance~\citep{hamming}, which measures the distance between two vectors based on the number of elements in common. Figures~\ref{fig:allsky_hamming} and \ref{fig:andromeda_hamming} illustrate the on-sky separation of all possible pairs of 10,000 randomly-sampled coordinates as a function of their Hamming distance, which in this case corresponds to the number of words different between each coordinate pair's \textsc{StarHash} names.

As a more stringent test against any local structure, Figure~\ref{fig:andromeda_hamming} densely samples a narrower sky area, specifically draws from Normal distribution, centred on the core of the Andromeda Galaxy, with a 0.5 degree standard deviation, though any sky location could be used without loss of generality.
In both cases, the desired behaviour of \textsc{StarHash} is recovered -- there is no correlation between the number of words in common between two \textsc{StarHash} names and their separation on the celestial sphere. Indeed, even for positions with an on-sky separation close to zero, a significant number of pairs of Hamming distance 2 and 3  are visible. The raw counts on each plot also highlight the core utility of \textsc{StarHash} -- with the lower Hamming distance bins having several orders of magnitude fewer pairs in them, meaning that the majority of identifiers are entirely distinct. Owing to the proprietary nature of similar existing geocoding algorithms, a direct quantitative comparison against \textsc{StarHash} is not possible, but there at least appears to be no statistically significant deviations from the expected separation.

Hamming distance is, naturally, only one of a number of metrics to measure the distance between strings. The Levenshtein edit distance~\citep{Levenshtein1966} is another such metric, which measures the number of single-character edits required to transform one string into another. Figure~\ref{fig:levenshtein_bonus} is included in the Appendix for completeness, amounting to a more granular version of Figures~\ref{fig:allsky_hamming} and \ref{fig:andromeda_hamming}.

\begin{figure}
    \centering
    \includegraphics[width=\linewidth]{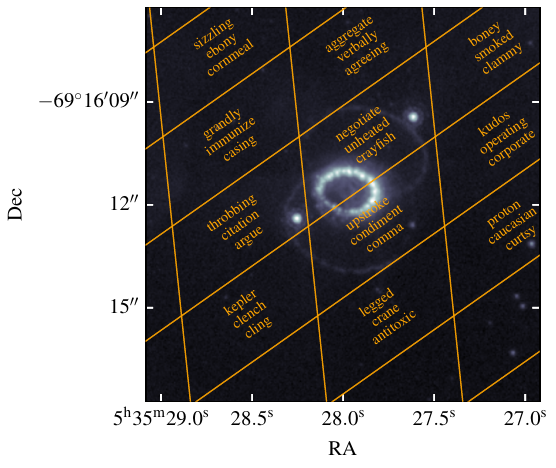}
    \caption{HST image of SN\,1987A, with the \textsc{StarHash} grid and names overplotted. The Author declines to comment on the academic suitability of some of the names. For visualisation purposes, the words have been stacked to provide a more compact representation over their hyphen-delimited form.}
    \label{fig:sn1987a}
\end{figure}

Whilst the memorability of a given word/phrase is hard to quantify, as it partly relies on human factors, I nevertheless show some curated examples, and leave it up to the Reader to decide. Figure~\ref{fig:sn1987a} illustrates the broad utility of \textsc{StarHash} in specifying regions of a given astronomical object -- in this case SN\,1987A. 
Whilst the centroid of the supernova lies in one grid square, the extended spatial structure spans multiple HEALPix pixels, with the grid size being well matched to smaller astronomical objects. This also highlights the robust spatial scrambling in a more qualitative sense.

\begin{figure*}
    \centering
    \includegraphics[width=\linewidth]{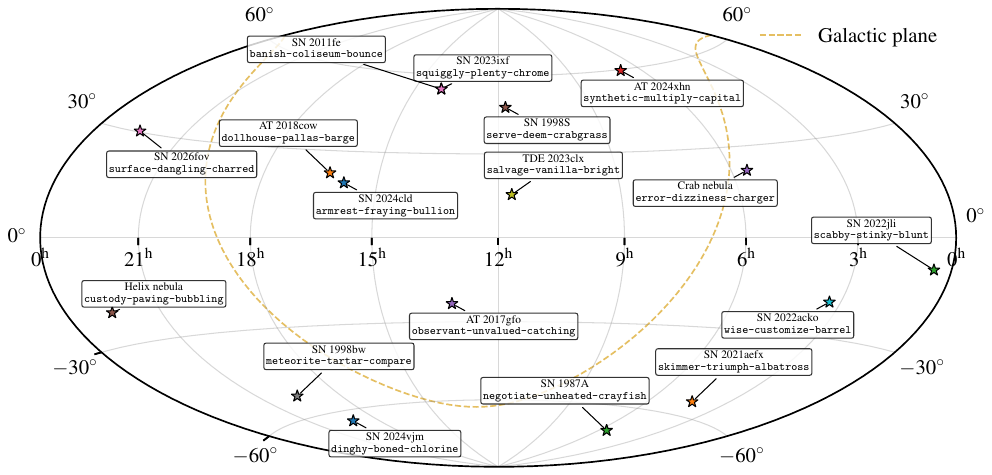}
    \caption{All-sky Aitoff projection showing the \textsc{StarHash} names for a selection of notable astrophysical transients\, alongside their conventional IAU-allocated names. The dashed line indicates the Galactic plane. These names may be decoded back to their respective celestial coordinates in a fully deterministic way following the methods in Section~\ref{sec:methods}.}
    \label{fig:notabletransients}
\end{figure*}

Figure~\ref{fig:notabletransients} provides an overview of the \textsc{StarHash} names for a number of important (highly-cited) and unimportant (worked on by the Author) objects\footnote{Unapologetically curated by the Author one random Wednesday afternoon to fill the available space.}, to illustrate how the ``new'' names are more memorable in many cases.

To assess the performance of the \textsc{StarHash} algorithm, as well as provide useful migration utilities for a number of catalogues, I generate pre-compiled name mappings for the IAU Transient Name Server catalog, and NASA's Exoplanet Archive~\citep{Christiansen2025}. On commodity hardware\footnote{The Author's laptop} \textsc{StarHash} achieves throughputs of $\approx6600$ names per second with only one thread. At this throughput, without multiprocessing or parallelisation, the entire Gaia DR3~\citep{GaiaCollaboration2022} source catalog may be \textsc{StarHash}-ed in just a few days, and would require only 11 GB of storage space to store the additional identifiers. \textsc{StarHash} is embarrassingly parallel, and benefits from near-linear scaling to significantly reduce the time to generate names for large catalogues. A full characterisation of the quality of randomness of \textsc{StarHash} names, their spatial correlation, and inevitable quirks is deferred to a follow-up publication, along with pre-computed mappings for larger catalogues.

\section{Discussion}
The reference implementation of \textsc{StarHash} introduced above will naturally surface some questions from the reader -- some of which I aim to address pre-emptively here.

A significant weakness of the \textsc{StarHash} naming scheme stems from its Anglocentric word list. Astronomy is an international science, and thus a naming scheme built on the English language disadvantages non-native speakers. Furthermore, words that are easily distinguished phonetically and semantically in English may be confused, or carry alternate (potentially unfortunate) meanings in other languages. Localised versions of \textsc{StarHash}, with region or language-appropriate wordlists would be one way to mitigate this, but introduce a separate issue -- the naming scheme is no longer universal. Whilst not intractable, such a problem clearly requires community consultation and effort to address.

The choice of 3.2 arcsec pixels is acceptable for optical time-domain astronomy, but woefully insufficient for high precision work such as radio astronomy, precision astrometry, or optical interferometry. Adding additional words from the existing wordlist (increasing $k$) is the most efficient way to ameliorate this in terms of scaling laws, at the cost of the brevity of the \textsc{StarHash} phrase. To get down to the maximum supported HEALPix resolution (0.4\,mas, $N_{side} = 2^{29}$), whilst keeping the current wordlist, would require only 5 words to be used, which is not overly prohibitive. To be able to specify the precise on-sky location of the shadow of M87* as seen by the Event Horizon Telescope~\citep{EventHorizonTelescopeCollaboration2019} would require a HEALPix grid of $N_{side} = 2^{36}$, with resolution of $\approx\mathrm{30\mu as}$. Such a grid can theoretically be covered with just one additional word (i.e. six words), a testament to the power of combinatorics. Knowing the Author's luck, the centroid would almost certainly fall on a HEALPix boundary.

The base reference implementation of \textsc{StarHash} presented here concerns itself solely with the location of an astronomical source, not the epoch at which it was discovered or observed. Whilst a design decision to simplify the rollout, this clearly presents a modest downgrade over existing naming schemes. Extending the hash space to encode time is an extension left to future versions, as decisions about the required time resolution must be made by the broader astronomical community. Nevertheless, adding a single word to the existing \textsc{StarHash} implementation to replicate the discovery year present in existing TNS/internal survey names would give sufficient hash-space for the next $\approx7000$ years of transient discoveries, which is more than satisfactory given the typical length of an astronomical career. Future generations of astronomers can consider extensions to the underlying \textsc{StarHash} standard at this point, assuming it hasn't been usurped by a superior naming scheme.

A significant bottleneck to the adoption of new standards is integration -- actually folding them into existing systems and deploying them. Whilst the \textsc{StarHash} codebase has been specifically developed to be production-ready for integration into existing transient brokers, survey pipelines, and side projects, barriers remain. Providing a hosted API service that provides an authoritative source of \textsc{StarHash} names and transforms between right ascension/declination and hash-phrases could significantly reduce the friction of adopting this new naming scheme. This is of course dependent on the availability of funds -- something fresh in the mind at the time of writing. A further planned enhancement, in defiance of conventional wisdom on premature optimisation~\citep{KnuthOptimisation}, is a complete rewrite of \textsc{StarHash} from the ground up in the Rust programming language. This would ensure full type and memory safety, and ``fearless concurrency''\footnote{\url{https://doc.rust-lang.org/book/ch16-00-concurrency.html}}, necessary for scaling to even greater volumes of discoveries in future, and helping the Author to obtain more transferable skills.

\section{Conclusions}
In this manuscript, I have presented \textsc{StarHash}, an algorithm for generating memorable, deterministic names for astronomical objects based on their coordinates. Building on the extensive literature for geohashing and prior art, \textsc{StarHash} represents a re-implementation for the demanding needs of astronomy: using HEALPix for segmentation, FF3 encryption for scrambling, and a curated wordlist to maximise the utility of the final compound identifiers. The statistical properties of \textsc{StarHash} have been validated, and example catalogs made available for the Reader to scrutinise and locate their favourite objects within.

\textsc{StarHash} is proposed not as the final word on astronomical object naming, but as a starting point for a broader, and genuine, conversation about the state of this crucial area of astro-nomenclature. With this in mind, the reference \textsc{StarHash} implementation and pre-computed catalog mappings are open source at \url{https://github.com/tkillestein/starhash}, and the Python package is directly installable via \texttt{pip}. Collaboration, dissent, and pull requests are welcomed.
\section*{Acknowledgments}
Joe Lyman and Lisa Kelsey are thanked for their fruitful discussions during the drafting of this paper. 
TLK acknowledges support via a Warwick Astrophysics prize post-doctoral fellowship, made possible thanks to a generous philanthropic donation. 

This research has made use of the NASA/IPAC Extragalactic Database, which is funded by the National Aeronautics and Space Administration and operated by the California Institute of Technology.
This research has made use of the NASA Exoplanet Archive, which is operated by the California Institute of Technology, under contract with the National Aeronautics and Space Administration under the Exoplanet Exploration Program.
This research is based on observations made with the NASA/ESA Hubble Space Telescope obtained from the Space Telescope Science Institute, which is operated by the Association of Universities for Research in Astronomy, Inc., under NASA contract NAS 5–26555. These observations are associated with program(s) 13405.
\bibliographystyle{apsrev4-1}
\bibliography{references}
\begin{appendix}
\section{Additional figures}
\vspace{10mm}
    \begin{figure}[h!]
        \centering
        \includegraphics[width=0.49\linewidth]{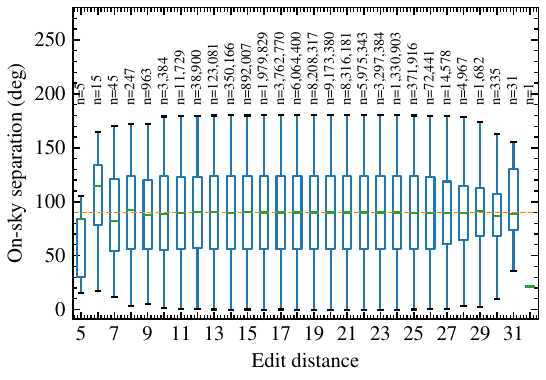}
        \includegraphics[width=0.49\linewidth]{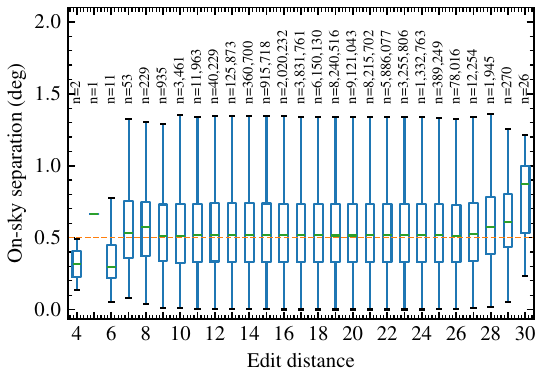}
        \caption{Box plot of Levenshtein edit distances of $\approx$50M unique coordinate pairs uniformly distributed across the sky (left) and drawn from a Normal distribution of 0.5 degree standard deviation (right) centred on the Andromeda Galaxy, with respect to their on-sky separations. The deviations seen at small and large edit distance are driven entirely by small number statistics and should not be interpreted as significant deviations from the rest of the distribution.}
        \label{fig:levenshtein_bonus}
    \end{figure}
\end{appendix}
\end{document}